# Mass data exploration in oncology: an information synthesis approach

Julie Bourbeillon[1,2], Catherine Garbay, Françoise Giroud[1]

**Abstract.** New technologies and equipment allow for mass treatment of samples and research teams share acquired data on an always larger scale. In this context scientists are facing a major data exploitation problem. More precisely, using these data sets through data mining tools or introducing them in a classical experimental approach require a preliminary understanding of the information space, in order to direct the process. But acquiring this grasp on the data is a complex activity, which is seldom supported by current software tools.

The goal of this paper is to introduce a solution to this scientific data grasp problem. Illustrated in the Tissue MicroArrays application domain, the proposal is based on the synthesis notion, which is inspired by Information Retrieval paradigms. The envisioned synthesis model gives a central role to the study the researcher wants to conduct, through the task notion. It allows for the implementation of a task-oriented Information Retrieval prototype system. Cases studies and user studies were used to validate this prototype system. It opens interesting prospects for the extension of the model or extensions towards other application domains.

[1] CNRS - Grenoble Universités, UMR 5525, Laboratoire TIMC-IMAG (Techniques de l'Ingénierie Médicale et de la Complexité - Informatique, Mathématiques et Applications de Grenoble). F38710 La Tronche, France

[2] CNRS - INRIA - Grenoble Universités, UMR 5217, LIG (Laboratoire d'Informatique de Grenoble). F38041 Grenoble, France

# 1 INTRODUCTION

Experiments in sciences are becoming more and more expensive in matters of time, material, equipment, etc. Two solutions have arisen among scientists in the biomedical field. The first one is the miniaturisation of samples and the automation of processes. A typical example could be the Tissue MicroArrays (TMA) technology, which allows for the mass treatment of hundreds of microsamples on a single histological slide. The second option is to reuse the data acquired by other teams, which are increasingly put on-line in an effort of resources sharing.

These approaches lead to a tremendous increase in the volume of available data. However these large masses of data also pose a real problem of understanding the data sets. This preliminary understanding is a mandatory stage for a more advanced exploitation. For instance, data mining tools have to be directed, which requires a minimal knowledge of the data space. In the same trend, hypothesis validation on an extract of a data set implies checking if the available information is sufficient. In the perspective considered in this paper, this process requires solving a set of complex problems:

- search and extraction of interesting data for a particular study, using potentially multiple data sources;
- aggregation of interesting items in a single information pool;
- organisation of relevant elements into a structure that facilitates interpretation;
- display of the relevant elements and of their structural organisation.

The complexity of these problems leads to an increasing need for a computerised assistance. The proposed solution is a synthesis notion, which federates the activities underlining the data grasping problem. It brings an original point of view on Information Retrieval (IR) by considering it as a component of the experimental approach.

This paper introduces this concept of synthesis in a particular application domain: the exploitation of data acquired during Tissue MicroArrays experiments. This application domain is presented in Section 2. After a state of the art review in Section 3, we present in Section 4 the model which supports our proposed theoretical framework. Section 5 introduces the resulting system architecture and the current prototype, and Section 6 presents its application in the TMA domain and a first validation of the approach.

# 2 THE TISSUE MICROARRAYS APPLICATION DOMAIN

## 2.1 The oncology research context

*In situ* molecular expression studies conducted as part of oncology research are a typical example of the information explosion phenomenon in the biomedical field. Indeed, oncology research relies, among other approaches, on the discovery of the oncogenesis mechanisms through tissue studies. The acquisition of such anatomopathological data classically relies on the construction, based on archived tissue samples, of histological slides on which the expression of interesting molecules is revealed.

Such processes fit well in the classical frame of the experimental approach, where hypotheses relating to a specific question are tested through experiments conducted on a limited group of individuals. But these processes are long to conduct, costly in reactants and lead to the depletion of non-renewable resources: tissue samples at research disposal.

## 2.2 TMA technology: to go beyond the limits of current approaches

In order to bypass these limitations, the TMA technology [1] collects and aggregates miniaturised biological samples into a gridlike receiver block, to support the high-throughput investigation of some phenomenon at the macroscopic level (e.g. evolution of cancers).

The overall process is presented in Fig. 1. Patients' samples are selected depending on the study to perform and organised in a TMA plan design. Small tissue cores (0.6–2 mm in diameter) are extracted from the corresponding biopsy paraffin blocks (donor block), then inserted into a TMA receiver block according to the plan. Slides are cut out of the TMA block and treated as conventional histological slides. Images of these TMA slides are then acquired and partitioned into individual spot images, where each spot corresponds to the slicing of one tissue core. Spot images are annotated by a pathologist; quantitative descriptors are further computed to characterise e.g. the staining intensity or marker distribution inside the cell compartments. The data sets consisting of images, annotations and quantitative evaluations are then stored in a database for further use. For instance data analyses are performed in the hope of getting answers or leads regarding the original question.

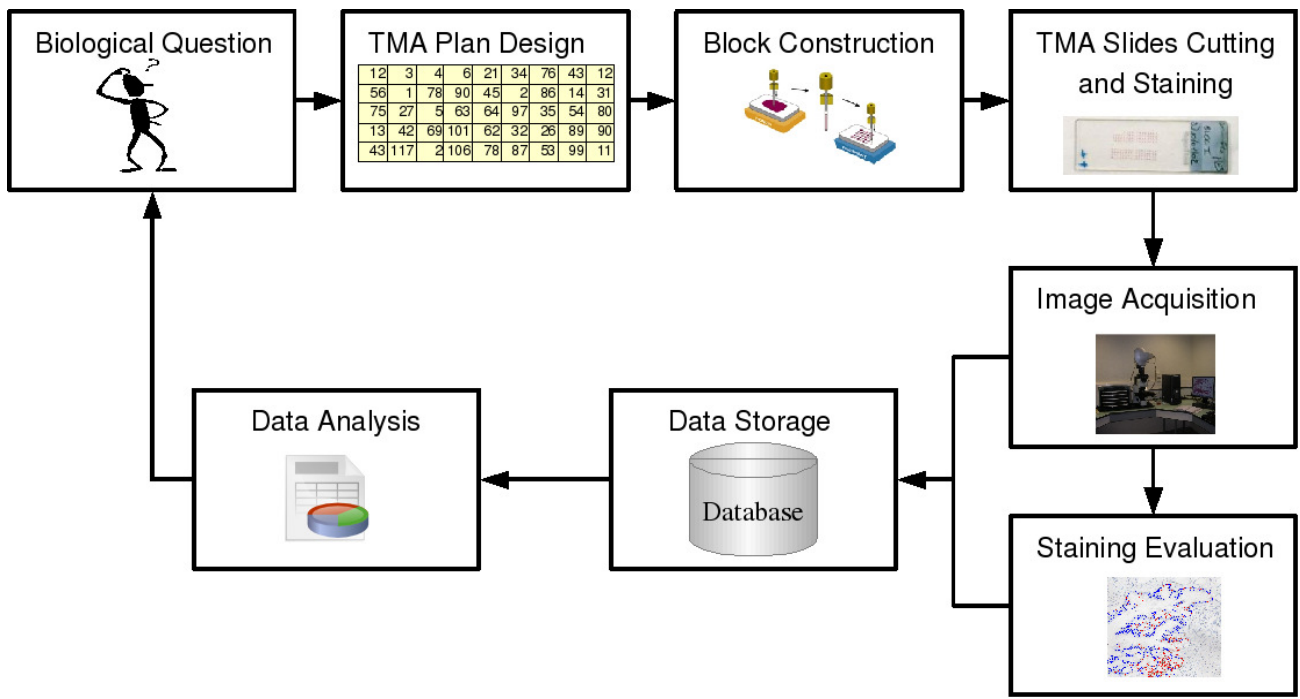

*Fig. 1. TMA technology – patients are selected according to the biological question addressed and corresponding core samples to extract are organised in a TMA design plan.*

### 2.3 New limits imply new approaches

This technology helps solving issues linked with the "whole slices" approach but its principle also leads to new limitations. For instance [2] proposes a method to solve the sampling issue and obtain a statistical representation of the whole biopsies. However the main problems are linked with the large volume of data. Indeed the high-throughput nature of the approach suggests a massive sharing of data. It also implies that the careful design of TMA blocks to test a narrow hypothesis is seldom conducted.

The large volume of data issue has been addressed through computer-driven data management. Most software focus on storage and data access [3], coupled with data visualisation [4] and image analysis or data mining tools [5]. But the biologist is still faced with a tremendous amount of information that is only loosely coupled with a biological question. Therefore the data grasping problem remains. One solution is the concept of virtual TMA slides [6] which displays data from several experiments at once. However the inclusion of relevant spots into the virtual slide is a manual process. Another approach goes towards careful design of TMA blocks according to a question, as presented in [7], but this is not always practical, e.g. when very few samples relevant to a problem are available.

Going beyond the limits of current tools when it comes to assisting scientists in a preliminary data exploration implies:
– providing assistance in hypothesis formulation to express the needs of a biological study;
– constructing a relevant data set on which the hypothesis could be tested;
– displaying this data set so that leads regarding the biological problem can be inferred;
– taking into account existing knowledge in the TMA domain and quality constraints.

The characteristics of this problem suggest it can be considered as a synthesis problem. Therefore we should explore the relevant concepts in the various domains which relate to synthesis.

# 3  STATE OF THE ART

## 3.1 Information synthesis: a multi-dimensional notion

In the scientific domain, synthesis appears mainly as an intellectual operation which consists in methodically reuniting the composing items of a whole. It emerges from the increasing need to compact a tremendous amount of information. This is achieved for instance through "review" papers or bibliographic syntheses, as introduced in the practical guide presented in [8]. Adopting this point of view on synthesis, we are faced with a concept with multiple facets such as mining data, retrieving information or representing parts of complex processes.

## 3.2  A data mining dimension

As the construction of a compact view on a large data set, synthesis can be linked with the data mining field [9], which aims at adding value to large data sets by extracting knowledge through mathematical and computerised methods. Several TMA-related software suits include such data mining packages [5]. However a successful use of these tools implies a background knowledge of both the underlying algorithms and information space to study [10]. In this context, tools allowing for the acquisition of a preliminary understanding of the data sets make sense.

The level of complexity involved in large data sets implies focusing on organisational and presentational issues. Information Visualisation [11] exploits the graphical abilities of modern computers to construct global views from data. Several tools providing rather compact and informative representations of data have been successfully used in the biomedical domain. For instance, Chan et al. [12] rely on growing self-organising maps to solve the binning problem posed by environmental whole genome shot-gun sequencing. Baehrecke et al. [13] uses Treemaps to visualise DNA MicroArrays data in relation with Gene Ontology annotations.

However the spacial organisation remains centred on attributes explicitly present in the data. Moreover the tools which allow for a focus on part of the data or a reorganisation generally consist in *a posteriori* treatments such as filters; synthesis implies the construction of views oriented a priori towards a precise goal. The problem is to select both the relevant information and tools, in order to display the appropriate view, given the objective.

## 3.3  An IR dimension

These selection problems are linked with the IR dimension associated with the synthesis concept.

The implementation of IR theories and algorithms [14] initially resulted in computerised systems relying on keywords. Pubmed[1] is a typical example in the biomedical field. However, while this model allows for evaluations as part of campaigns like TREC[1], it may not reflect the complexity underlying information access. Example solutions could be enhancements added to Pubmed[2], such as improved document indexing approaches [15] or the inclusion of visualisation techniques to present the results [16].

The most interesting approaches might be those aiming to satisfy the need of the user, which is the focus of information behaviour research. The user is considered as a thinking entity in a context which can be socio-economic, cultural or affective [17]; he is faced with the cognitive problem of an incomplete representation of the world. The IR process therefore aims at extending his knowledge. These studies are mostly theoretical or based on experimental studies of users. For instance, Grefsheim and Rankin [18] survey information needs and search behaviour of researchers and research administrators of the US National Institute of Health. This type of studies led to models of the interaction mechanisms between a user and information sources [19]. Unfortunately they cannot directly lead to IR implementations.

The need for operational systems raises the issue of the use which is made of the retrieved information. This problem is pointed out in the field of bioinformatics in [20]: IR is integrated as part of an experimental protocol implying progressive hypothesis refinement steps. Introducing the notion of problem solving or task in the IR process appears to be a major issue [21]. This dimension still remains underexploited or implicit, because of the complexity of its understanding, analysis and expression.

---

[1] http://trec.nist.gov/.
[2] http://www.ncbi.nlm.nih.gov/pubmed/

Among the approaches in this direction, Choo et al. [22] aims at designing taxonomies of tasks, by collecting data regarding the behaviour and motives of user groups. These taxonomies can be used to expand queries with user-selected and task-oriented controlled vocabularies [23]. But they remain difficult to exploit because they are too general [22] or limited to one particular task [20].

### 3.4 A knowledge representation dimension

These various attempts at enriching the search context have a common feature: they rely on a representation of knowledge regarding the user or the task he performs. A deeper focus on the user is provided by adaptive hypermedia [24], which constructs a representation aimed at a specific audience. To achieve this goal, a user representation is devised to offer personalised services [25]. Such approaches are used for instance in medical tutoring systems to adapt exercises to the student [26].

The task notion is central to some Artificial Intelligence paradigms such as Problem Solving Methods and knowledge representations. These can be used to enrich the retrieval context. For example, the medical problems-solving steps in the tutoring system from [27] are represented using the Unified Problem-Solving Method Description Language. Nowadays the objects manipulated through these tasks are represented through ontologies [28]. Numerous examples exist in the medical domain [29], from simple controlled vocabularies such as Gene Ontology[1], to complex representations in descriptive logics such as the Foundational Model of Anatomy[2].

## 4 SYNTHESIS MODEL

We are proposing a system focused on information synthesis tasks, supported by a grid-based display, which is inspired from the organisation of items on a TMA slide. This particular viewpoint implies reconsidering the three steps of query formulation, information retrieval and information display, as sketched in Fig. 2.

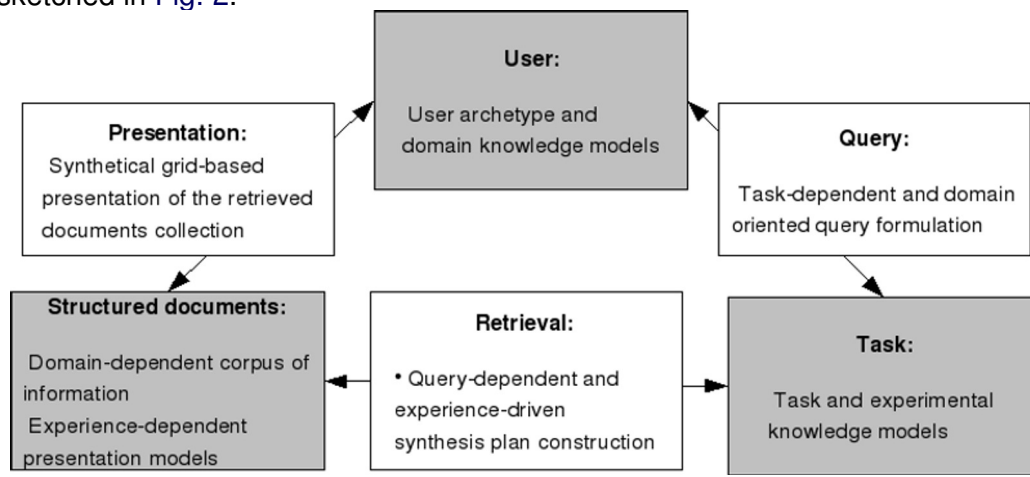

*Fig. 2. Synthesis model – the synthesis process results from the interaction between three entities: user, task and structured documents. It implies several interactions: task*

Query formulation is extended from the specification of keywords to include some task specification. Information retrieval is extended from the matching between keywords and document to involve the elaboration of a plan to select and combine the appropriate information. Information display implies the composition of a complex multimedia document instead of a mere list.

### 4.1 The task level

Some general task taxonomies have been devised through information need studies [19,22]. While they provide an overview of the wide range of tasks to be considered, they give rather few ideas on how to make task expression operational. We focus on information synthesis tasks [30]. In all application domains, these tasks may be seen as a finite hierarchy of prototypical tasks. A first attempt at a taxonomy is presented Fig. 3.

---
[1] http://www.geneontology.org/.
[2] http://sig.biostr.washington.edu/projects/fm/.

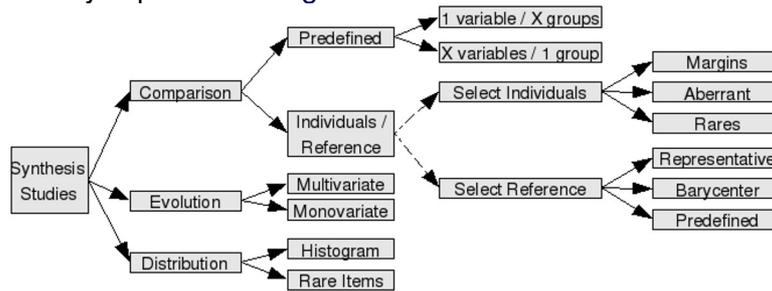

*Fig. 3. Task taxonomy – synthesis tasks are considered as belonging to three categories which can be further divided into finer grain sub-categories: comparisons provide insight into the structure of the data space, evolutions provide an overview on the correlations between items and distributions provide access to the repartition of the individuals. This hierarchy is represented by solid arrows. Sub-categories can also constructed by the combination of leaf items from different branches represented by dash arrows. For instance, a comparison of individuals towards a reference can be obtained by comparing aberrant individuals towards the barycenter of the group.*

Being IR tasks, they imply the usual selection and presentation steps. Being synthesis tasks, they imply an organisational step, to support the critical examination of the selected items. Each task is therefore modelled as the composition of three major sub-tasks (selection, organisation and presentation), which themselves are complex and can be decomposed as sub-task trees. These sub-task trees are therefore orthogonal to the previous taxonomy and define the Task Model corresponding to each task. A simplified view of a "Comparison" Task Model is presented in Fig. 4.

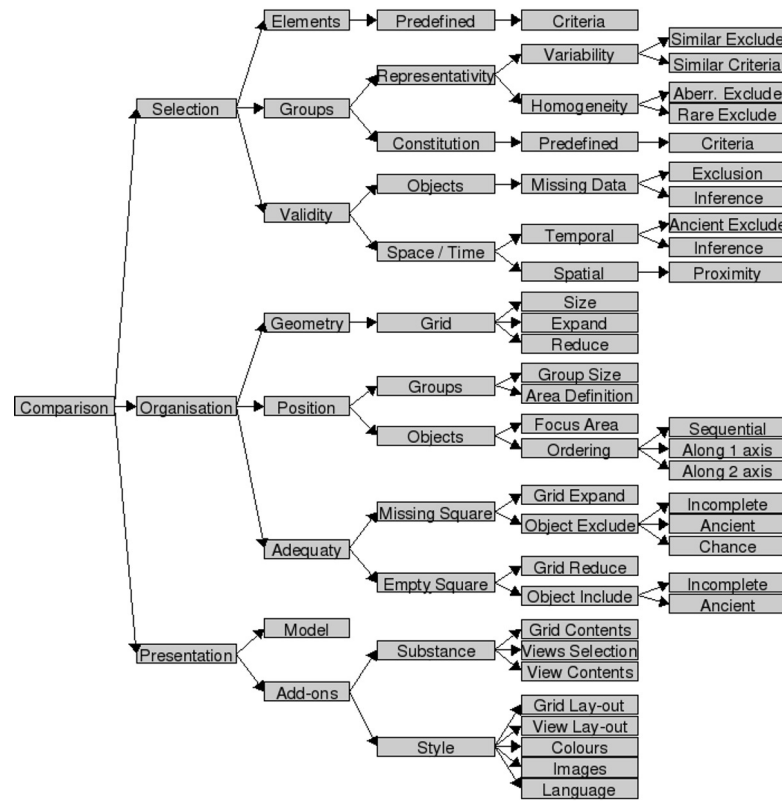

*Fig. 4. Simplified Task Model – a ''Comparison'' task is decomposed in a sub-tasks tree. Each rectangular node represents a sub-task category which is decomposed at the next level until leaves are reached. These leaves represent elementary sub-tasks.*

Synthesis tasks are further influenced by some application domain knowledge which we call experimental. It is heterogeneous and ill defined; it uses heuristics to drive the practical exploration of data. For instance, the "Comparison" Task Model integrates various heuristics defining ways to handle missing information: "Exclusion", which discards items with missing data, or "Inference", which computes missing values using similar items. The effective heuristics to be applied are selected dynamically, at run-time, according to experimental knowledge.

### *4.2 User archetype and query formulation*

Query formulation by users strongly depends on their knowledge of both the IR system and the application domain. We resort to user archetype models, which represent a class of users sharing specific domain knowledge and a common way of working in this domain. The existence of such mental models has been studied among attorneys [31]. In our approach this archetype includes a taxonomy of the domain vocabulary and a view of Experimental Domain. These archetypes are further individualised by preferences.

Including the notion of task alters the nature of the query. It becomes a n-tuple where each tuple plays a different role towards the task. This implies using Query Models such as the one which is schematically presented in Table 1. Formulating a query then consists in selecting a domain, choosing a task of interest to this domain, and finally specialising the corresponding Query Model.

*Table 1. Example of a Query Model for a comparison task.*

| ***Model Item*** | ***Description*** |
|---|---|
| **Generalities** | |
| Task | Task Category, e.g. Comparison |
| Title | Short description of the synthesis goal |
| Description | More precise description of the study |
| Domain | Application domain the query is related to |
| **Needs** | |
| Goal | Target item, e.g. item to compare |
| Inclusion Criteria | Criteria guiding the selection of items (classical IR terms) |
| Organisation Criteria | Criteria guiding the organisation of the items to handle, e.g., criteria defining the composition of the groups to compare |
| Ordering Criteria | Criteria guiding the organisation of the items inside each group |
| **Experimental constraints** | |
| Language | Language to use for the display |
| Colour | Colour scheme to use inside the synthesis document to represent the values of the variables |
| Selection criteria application method | Approach to use in the application of the selection criteria, either strict (exact match) or approximate (similar values) |
| Missing data management | Approach to use to manage items with missing data, either exclude the items or infer missing data from similar items |

### *4.3 Structured documents*

Information visualisation techniques [32] must be used to support the structured presentation of the synthesis result. The usual ordered list is not sufficient. It is replaced by a complex grid-like structure, where each item is considered as situated in the context of other items, with which it may either share some properties, or conversely differ for others. Using a grid allows for a simple and compact visualisation of the search results; it offers a relational overview over the collection at hand, bringing into light major tendencies as well as unexpected phenomenons (regularities, breaks, commonalities or rare events.).

# 5 ARCHITECTURE AND PROTOTYPE

## 5.1 Architecture

The proposed model has guided the design of an architecture for a synthesis system prototype. Synthesis document generation is performed along a two-stage process by the engine presented in Fig. 5 [33]. At the heart of this architecture is the notion of Task Instance, which is built as the result of the query formulation process, and guides the elaboration of the synthesis document.

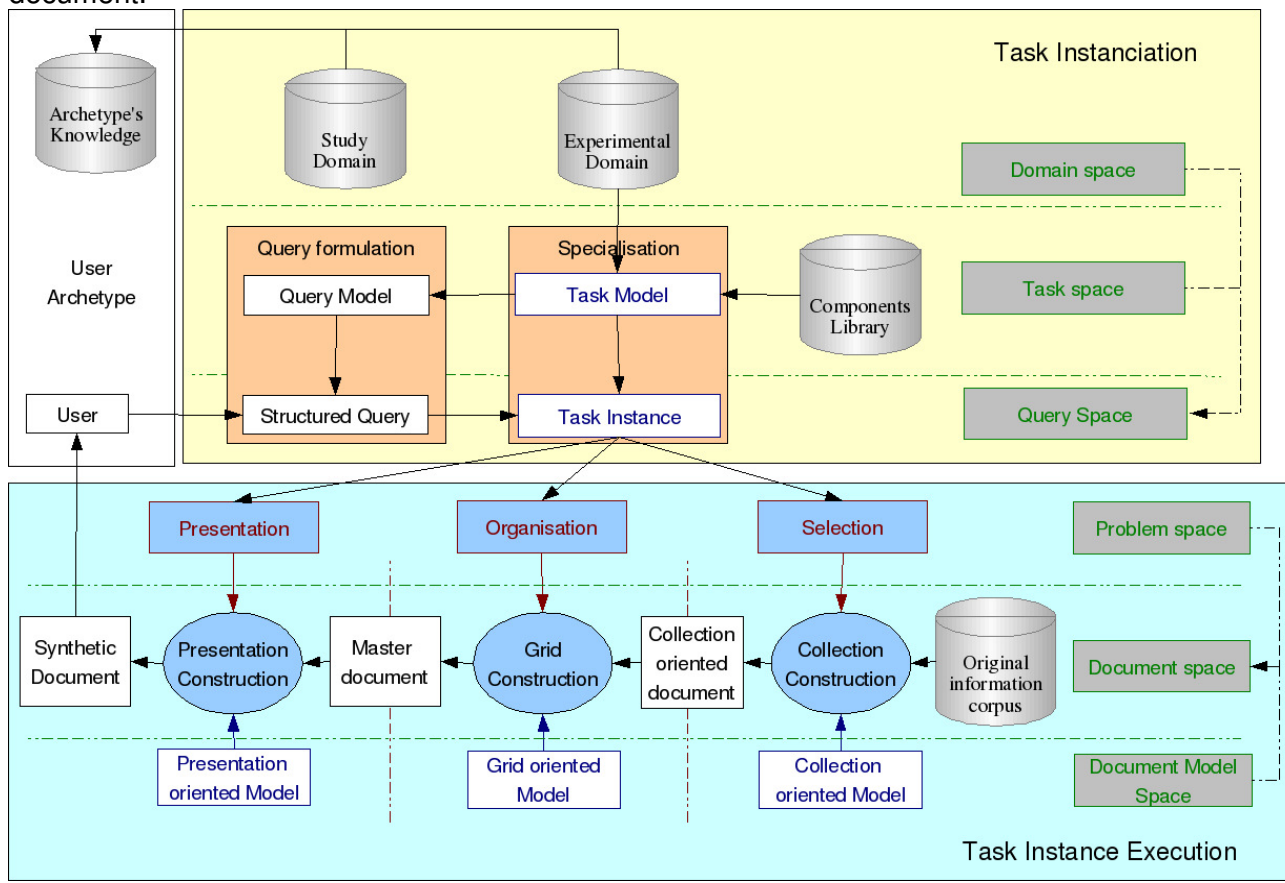

*Fig. 5. Architecture – the synthesis system consists in two main functions: a task instanciation and the execution of this Task Instance. These functions can be decomposed in finer grain operations and imply the manipulation of entities.*

This Task Instance is generated along several specialisation steps. The user first formulates a Structured Query. He specialises the Query Model corresponding to the task he is interested in (selected in the taxonomy from Fig. 3), using Study Domain and Experimental Domain terms corresponding to his Archetype. This Structured Query is then used to build a Task Instance, based on a generic Task Model, which specifies the main components supporting the successive selection, organisation and presentation steps (as presented in the sub-tasks tree from Fig. 4 for a comparison task).

The Task Instance is then executed. This is performed according to three stages, each corresponding to a major sub-task of the Task Model: Selection (a set of relevant documents is selected), Organisation (the spatial organisation of the selected items is specified) and Presentation (the lay-out of the final synthesis document is specified).

## 5.2 Prototype

A first prototype has been built to experiment the potential of the proposed notions. This prototype is developed in JAVA and relies heavily on XML: ad-hoc schemata accessed through the JAXB library for the domain knowledge, query, generated documents and configuration files ;

XFORMS (through the Orbeon Forms[1] framework) for the user interface. Each elementary subtask corresponds to a distinct software component which parameters are defined in the query.

### 5.2.1 Task instance construction phase

The Task Instance construction phase is performed according to the schema presented in Fig. 6.

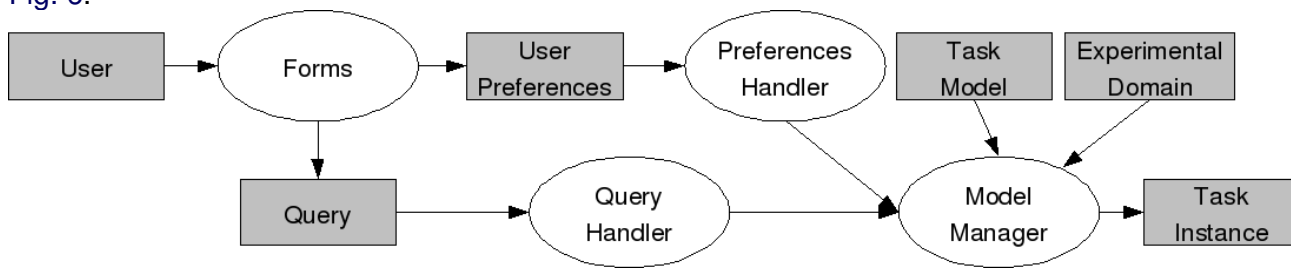

*Fig. 6. Software architecture for the elaboration of the Task Instance.*

Central to this stage is the Model Manager. This piece of software goes through the Task Model. It designs the Task Instance by integrating information from the Structured Query and User Preferences along with Experimental Domain knowledge. Its main objective is the selection of components and the fusion of information, to instantiate the task in a manner which will lead to the most relevant synthesis document.

This stage is difficult because several components may be available to solve the same sub-task (for example the "Missing data" sub-task). Heuristics stored as Experimental Domain knowledge should help solve this issue, but they may be insufficient to guarantee the overall relevance of the synthesis document. An additional problem comes from the necessity to fuse potentially heterogeneous and contradictory information extracted from various sources (Query, Preferences, Experimental Domain).

In the prototype both issues have been put aside. We forced the use of one single component for each sub-task. We also defined a precedence order between information sources. Further work will be necessary towards a fully autonomous prototype.

### 5.2.2 Task instance execution phase

Constructing the synthesis document implies solving each subtask included in the Task Instance according to the schema in Fig. 7.

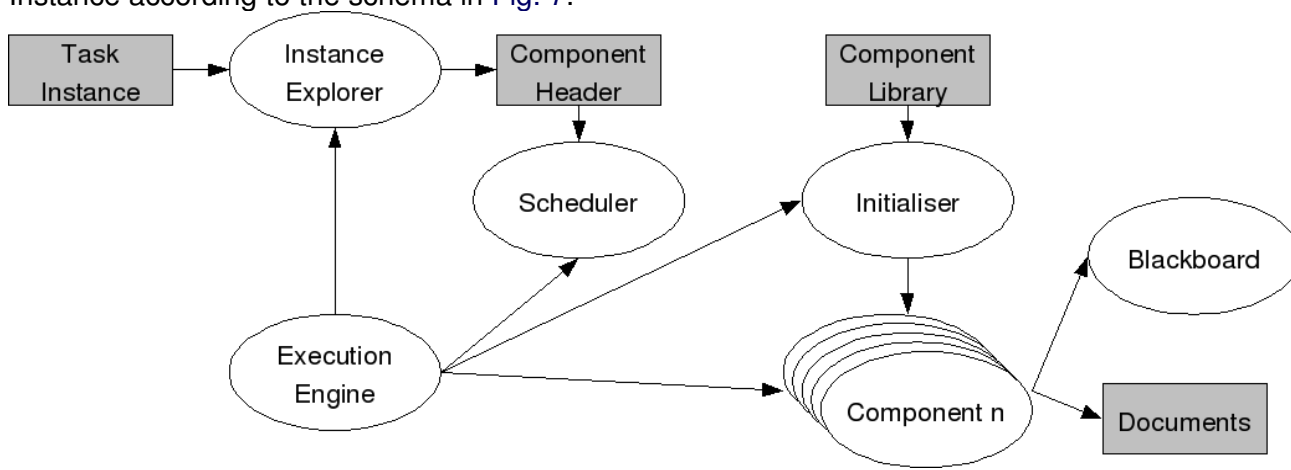

*Fig. 7. Software architecture for the execution of the Task Instance.*

This phase is performed by the Execution Engine. It reads the Task Instance thanks to an Instance Explorer which translates it into a set of Component Headers, which describe the Components' specialisation instructions along with their input/output descriptions. The Engine then resorts to a Scheduler to define the Components' initial running order. Once this order has been established, the Engine starts executing each Component in the specified order. Components are loaded from the Components Library, and then initialised and run according to the instructions of the Header.

---

[1] http://www.orbeon.com/.

This process poses two major issues. First of all, dependencies among components are not enough to deduce their running order: the dynamics of the document construction are complex and implicit dependencies play a major role. For instance, it might be required to re-run some components if their previous execution lead to a dead end. Moreover some sub-tasks are both very complex and impossible to decompose into a set of simpler logical sub-tasks. The components we have proposed so far to solve these sub-tasks are still rudimentary and will have to be refined.

# 6 EXPERIMENTAL VALIDATION

## 6.1 Validation goals and procedure

The evaluation of the current realisations can be considered along two axes. First of all the proposed system is a piece of software. Most software evaluation metrics have little relevance in the context of a prototype, except the evaluation of the use of the system. This suggests conducting case studies. Moreover the synthesis model is based on IR models. Objective IR measures such as relevance would imply taking into account the lay-out of the synthesis document. They are too complex at this stage. However user studies based on questionnaires can be used to capture the potential users' opinion.

In this section, we will present a twofold experimental validation, based on case studies and a preliminary user study.

## 6.2 Case studies

The preliminary case studies presented here are based on a limited data set so that the results can be manually checked. The cohort consists in 162 patients suffering from colon lesions followed at the Centre Régional de Lutte Contre le Cancer de Montpellier (France) who underwent surgery between 1988 and 2002. The median age of the patients is 66 (range, 24–91). They include 70 female, 90 male and two sex unknown. The localisation of their lesion represents the various areas of the colon, cancer stages are various and even if adenocarcinoma are the majority, other lesion types are represented. Data extracted from their clinical files and the evaluation of several staining measures by a pathologist on corresponding whole section and TMA slides lead to a database including 65 parameters per patient.

Given the development stage of the prototype we focus on two aspects of the approach:
- Query formulation: how may a natural language biological problem be expressed at a macroscopic level and translated into a n-tuple representing the Structured Query?
- Synthesis document: does this construction, which is visualised as a multimedia document, allow for knowledge inferences and is it really relevant for the user?

### 6.2.1 Comparison case

#### 6.2.1.1 Structured query

We decided to analyse whether there are significant differences between tumoral tissues and tissues adjacent to the tumour. Indeed the interface between the non-tumoral and tumoral tissues is not well known. We therefore analysed key molecular elements involved in colon carcinogenesis and compared their expressions in both tissue types:
- β-catenin: cellular adhesion molecule, involved at the beginning of tumour progression ; mitosis signals address it to the nucleus where it acts as transcriptional activator for various molecules, among which Cyclin D1;
- Cyclin D1: molecule regulating the cell cycle by contributing to mitosis initiation and cellular proliferation;
- Ki67: widely used marker of cell proliferation, preferentially expressed in the nucleus of active cells;
- Bcl2: apoptosis inhibitor.

We focused on the relationship between cellular localisation and expression of the markers in each tissue type. The biological problem therefore is: "comparison of the intracellular localisation of the expression of each marker in tumoral and adjacent tissues from all patients with colon carcinoma".

This query can be formulated as being a kind of "Comparison" task, which goal is to compare the markers' expression, as measured for example by "the percentage of marked cells". The focus of this study (i.e. inclusion criteria), here the organ under examination, is the "colon". Finally, the groups to be compared (i.e. organisation criteria) are selected as being tumoral and adjacent tissues (i.e. Diagnostic for the tissue), under the various possible marker intracellular localisations (i.e. Description for the cell).

This formulation leads to the representation in Fig. 8: the drop-down lists of the query interface allow to go through the domain taxonomy and express the needs of the user. The input system and the underlying representation therefore show a sufficient expression ability to define biological problems as complex as the one addressed here.

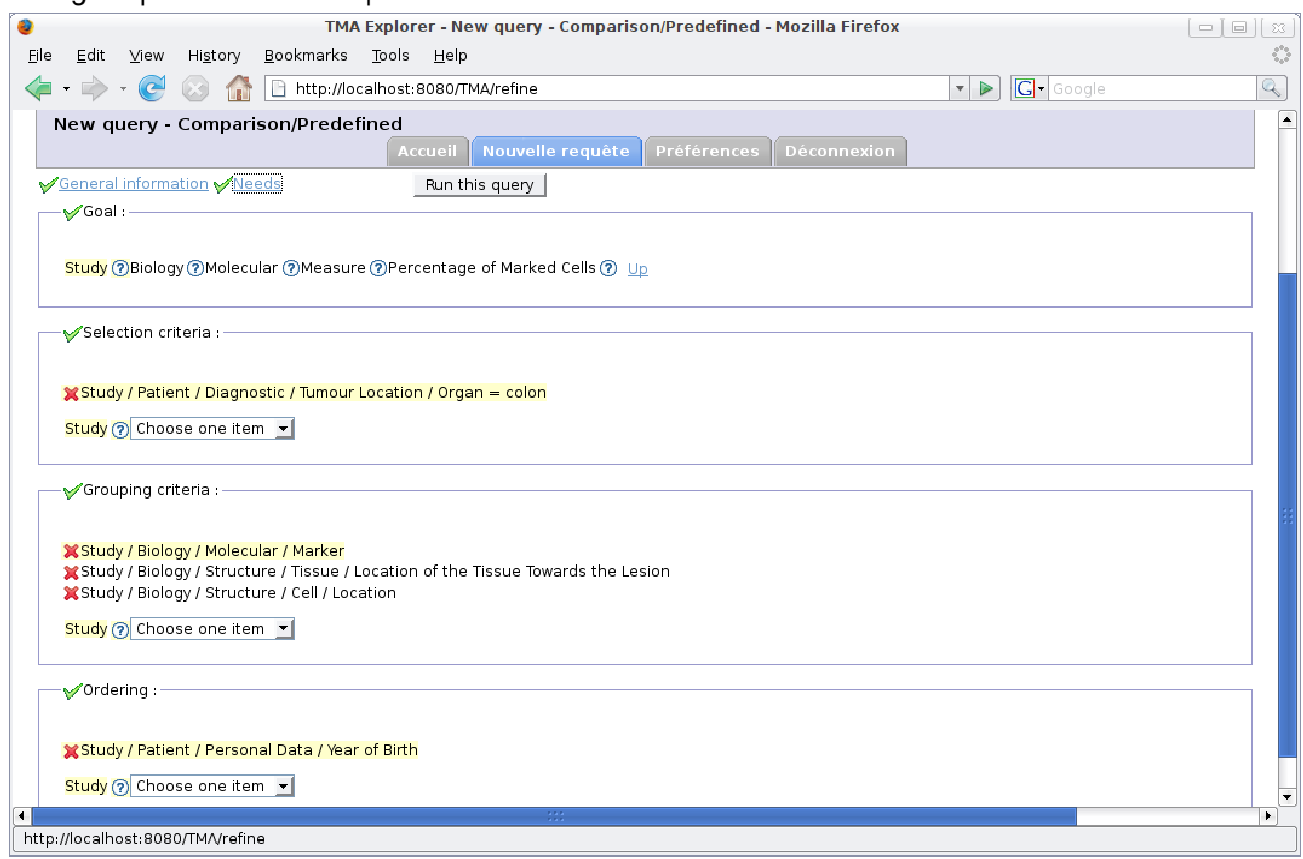

Fig. 8. A view of the query interface when filled to formulate the biological problem from the comparison use case: ''Comparison of the intracellular localisation of the expression of each marker in tumoral and adjacent tissues from all patients with colon carcinoma''.

### 6.2.1.2 Synthesis document

Some screenshots of a synthesis document are presented in Fig. 9. The documentary grid represents each individual item (TMA spot) as a square, which grey level corresponds to the percentage of marked cells (goal of the comparison), from 0 (white) to 100% (black). These spots are grouped by areas, which are delimited by titles describing their contents. Each spot also serves as a link to various relevant views of associated information: biopsy block the TMA core comes from, patient clinical data...

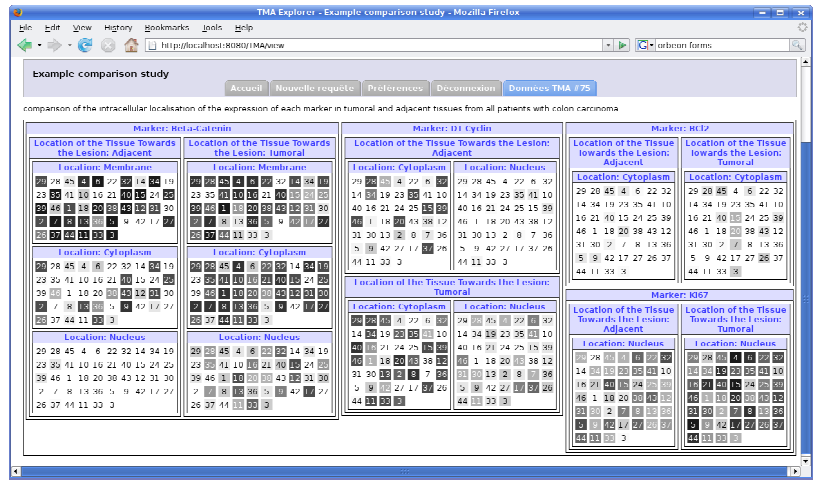

*Fig. 9. Synthesis document constructed as a response to the query presented in Fig. 8. Here is displayed the documentary grid. It represents each individual item as a square spot which grey level corresponds to the percentage of marked cells (goal of the comparison), with about a dozen of levels, from 0 (white) to 100% (black). These spots are grouped by areas which are delimited by titles describing their contents. Each spot also serves as a link to other items and gives access to various relevant views on associated information.*

Only 46 patients are displayed, because individuals with missing data were excluded by the system. Moreover, for each marker, the staining level depends on the considered tissue (tumoral or adjacent to the tumour) and the intracellular localisation:

- β-catenin (left part of the grid): for normal cells (left) the β-catenin localisation is primarily in the membrane, whereas for tumoral cells (right) the staining goes through the whole cell. However we know that, in resting normal cells, β-catenin is expressed in the membrane and then degraded in the cytoplasm. In tumoral cells, mutations and / or other upper biological events in the β-catenin pathway prevent this degradation. The β-catenin then migrates to the nucleus where it activates cyclins transcription.
- Cyclin D1 (middle): for normal cells (top) Cyclin D1 is mostly located in the cytoplasm, whereas for tumoral cells (bottom) it is also present in the nucleus and the staining is higher. We know Cyclin D1 contributes to mitosis initiation. Higher Cyclin D1 levels imply an increased mitosis rate and higher proliferation level, which fits with tumoral tissues where cells division is fast and anarchic
- Bcl2 (top right): staining is low but spread wider among patients in tumoral tissue (right) than in normal tissue (left). Furthermore we know that this molecule inhibits apoptosis. Our findings of increased Bcl2 staining in tumoral cells imply a reduced cellular death rate, which is to be expected,
- Ki67 (bottom right): the staining involves higher numbers of nuclei in tumoral tissues (right) than in normal tissue (left). This is to be expected, since tumours have an increased cellular growth pace.

The documentary grid of the synthesis document therefore offers an overall picture both simple and fitting with known biological facts. But it is however necessary to confirm these intuitions with a statistical analysis of the data. Therefore a Wilcoxon rank-sum test was used to compare the groups among tumoral and ajacent samples (data not shown). This test showed significant differences between tumoral and adjacent tissues, for all molecules, in all intracellular compartments ($|z|$ close to 0), except membrane β-catenin. This corresponds to the groups which were identified in the synthesis document. The synthesis paradigm can therefore be used as an exploratory method to help in orientating statistical analyses.

### 6.2.2 Evolution case

#### 6.2.2.1 Structured query

For colon carcinoma, the pTNM stage of the tumor is a central prognosis element. This classification includes three components, T, N and M, which each represent one dimension of the tumoral invasion: invasion of the different tissue layers for T ; invasion of the lymph nodes for N ; distant invasion through metastasis for M. Therefore, the higher the values of T, N and M, the worse the prognosis for the patient.

The question one can then ask is the independence between the various dimensions of the classification: "is there a relationship between the lymph node invasion and the tissue invasion?" Therefore the second study example will be the "evolution of the number of invaded nodes depending on the number of observed nodes with observation of the T component of the stage for patients with colon carcinoma".

In this particular example, the formulation of the query (see Fig. 10)from the description of the problem seems possible just as it was in the previous case.

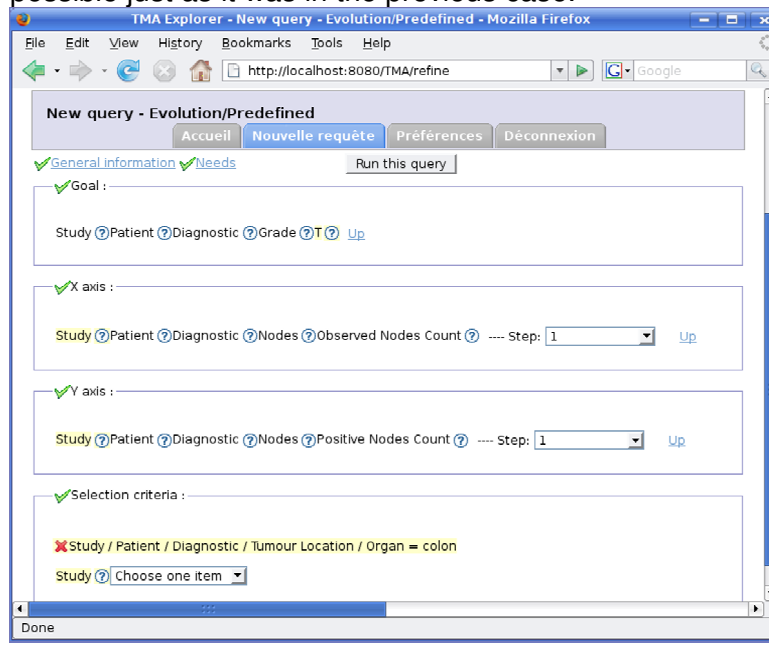

*Fig. 10. A view of the query interface when filled to formulate the biological problem from the evolution use case: ''evolution of the number of invaded nodes depending on the number of observed nodes with observation of the T component of the stage for patients with colon carcinoma".*

### 6.2.2.2 Synthesis document

A screenshot, corresponding to the documentary grid of the synthesis document, is presented Fig. 11. This grid organises pertinent items according to the two axes defined in the query: the number of observed nodes and the number of positive nodes. For each combination of values along these two axes, the average individual within the set is displayed. This leads to a representation with less displayed individuals (73) than patients in the database (162). Here, the average individual corresponds to the patient with the closest value to the average of the group for the goal of the study (here the T component of the stage). This individual is represented by a square whose background colour is linked to the goal of the study, from light grey (stage 1) to black (stage 4). The number in the square corresponds to the identifier of the individual and provides access to his information sheet.

This synthesis document can be used to observe the structure of the data set, for instance to identify aberrant data. Here we observe the number of positive nodes among the observed nodes. There should not be more positive nodes than observed nodes. But it is not the case for patient 3: it looks like a typical data entry error.
We also have to consider whether the document helps answering the biological question. To do so, we can compare patients with no node invasion (bottom row) to the others. The output does not show any difference regarding the T component of the stage between the two groups. We can therefore consider the two component T and N of the classification are independent, which is in fact known biological information.

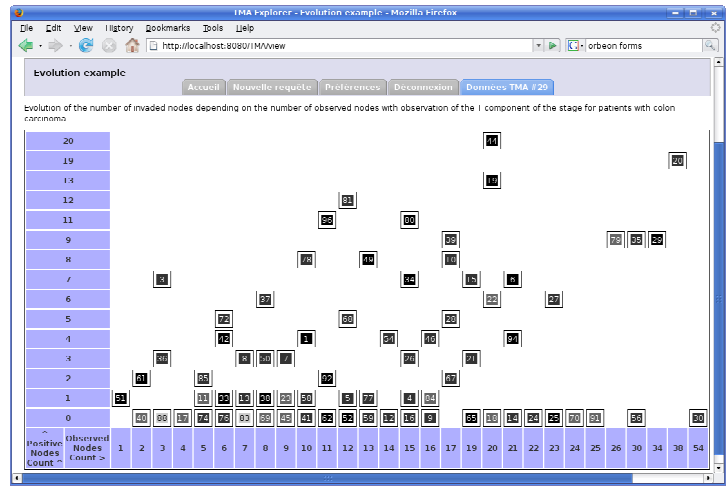

*Fig. 11. Synthesis document for the evolution example whose query is introduced in Fig. 10. For each combination of values along these two axes for which items exist the average individual within the set is chosen. This individual is represented by a square whose background colour is linked to the goal of the study (here the T component of the stage), from light grey (stage 1) to black (stage 4). The number in the square correspond to the identifier of the average individual and provides access to his information sheet.*

As for the comparison case, the intuition drawn from the synthesis document has to be statistically confirmed. Therefore a Pearson's chi-square test was used to evaluate a potential dependency between node invasion presence and T component of the stage (data not shown). This test showed the two are independent (p-value close to 0). The observation of the synthesis grid allowed to point out known biological facts, which suggests that this kind of evolution tasks could be used to evaluate new hypotheses.

### 6.3  User study

#### 6.3.1  Objectives

We are focused here on the potential users' point of view. Since it is difficult to devise objective metrics, the evaluation is based on answers to a questionnaire and users' comments.

The general goal is to evaluate the usability and efficiency of the system. Usability includes ergonomics, navigability, learning curve, etc. Efficiency can be considered along two axes: a "user" efficiency, which can be related to the user relevance of IR systems, and includes query formulation and result interpretation dimension ; a "system" efficiency, which corresponds to performances measures.

#### 6.3.2  Experimental settings

Given the limited time available to conduct the user studies the size of the user panel was reduced (8 users), which limits the possibility of statistical analyses of the results. However the group included a wide range of profiles, with variable combinations of expertise along two axes: computer literacy and oncology.

The tests were conducted according to a defined test scenario:

– Query formulation and result interpretation for the two biological problems from the case study,
– Query formulation and result interpretation for user defined biological problems,
– Query reformulation leading to a change of focus on the comparison case.

The associated questionnaire included 28 sentences for which the user had to provide an approbation level from 1 (disagree) to 5 (completely agree). These were organised according to three categories: "Generalities" covering usability issues, "Query formulation" and "Results interpretation". The observations of the user behaviour was conducted by an experimentalist who evaluated the time spent on each screen, measured the computing time and recorded the remarks spoken aloud by the users.

### 6.3.3 Results

#### 6.3.3.1 Usability

The "Generalities" section of the questionnaire is declined along several axes: ease of use, conviviality, ease of learning, navigability and overall satisfaction. The size of the panel being small, no definitive conclusion can be drawn, but in general the users seemed interested in the prototype they tested. The only restrictions are minor and have to be expected from a prototype: some ergonomics flaws or limited contextual help, etc.

In the context of an unusual display such as ours the ease of learning dimension is of particular interest. Indeed it is closely linked with the adaptation time, which is essential for the adoption of a new technology. It has been evaluated by measuring the time spent to formulate queries (see Fig. 12) and interpret results (data not shown).

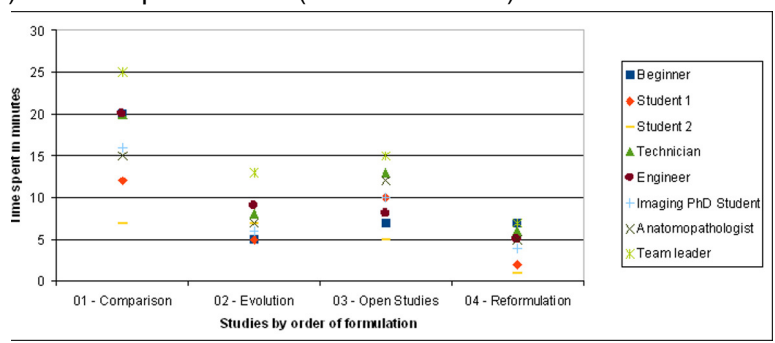

*Fig. 12. Time spent for query formulations by the various users depending on the study.*

These measures show a general decrease of the time spent over the various queries, except for the open studies. This can be explained by the time required to devise a new biological problem, whereas the problem is already defined for the other queries. There seems to be a significant effect of learning on the ease of use of the tool for both query formulation and synthesis document interpretation.

#### 6.3.3.2 Query formulation and synthesis document interpretation

The "Query formulation" section of the questionnaire is declined along several axes: adequacy of the query to the task at hand, actual formulation, completeness, expressiveness, extensibility and navigability. The results seem to indicate an overall satisfaction of the users. The only limits relate to the completeness and expressiveness. Regarding the completeness, users expressed a need for other tasks. The expressiveness of the query is limited by the lack of a boolean OR and the impossibility to build composed concepts ("the staining intensity for the marker Ki67" for instance). This should be solved by future developments.

The "Results interpretation" section of the questionnaire explores also several dimensions: intuitiveness, informativeness, utility, suggestiveness and navigability. The results are here again positive. The grid representation is considered as relevant, sensible and as bringing new information. The intuitiveness is the less positive axis: users have to get used to the representation as our analysis of the learning process has shown. Users also showed their enthusiasm through the amount of suggestions they have made, either to facilitate the interpretation of the synthesis document (adding more colour-coded elements, allow for a joint view of several documents, add exportation tools to ease reuse in publications) or to extend the document (add a display of simple descriptive statistics and links to data analysis tools, improve the interaction with the grid through zoom/focus or selections as basis for a new study).

#### 6.3.3.3 System performances

As part of the development of a prototype, an evaluation of the performances allows to point out possible flaws, in the perspective of developing a complete production system. This evaluation was conducted in an informal manner through rough estimates evaluated during the user study on a Pentium IV dual-core 1.83GHz with 1Go RAM. Table 5 presents the results.

*Table 2. Results of the performances evaluation.*

| Item | Minimum time | Maximum time |
|---|---|---|
| Display: Homepage | 1s | 3s |
| Display: Query – Generalities | 1s | 3s |
| Display: Query – Experimental constraints | 2s | 4s |
| Display: Query – Needs | 1s | 3s |
| Display: Query – Reformulation | 12s | 20s |
| Computation | 2s | 2min |
| Display: Synthesis grid | 2s | 10s |
| Display: Patient page | 3s | 5s |
| Display: Histology page | 3s | 6s |

For most items these measures indicate delays within a few seconds range which is compatible with an interactive system. However two items pose problems: computation of the synthesis document and display of reformulation forms. Both issues have been analysed and solutions individuated.

## CONCLUSION

In this paper we took an interest in defining a task oriented IR model, which led us to explore the notion of synthesis task. The construction of a prototype based on this model allowed for a first validation of the task-oriented IR concepts applied to the biomedical domain. The current prototype is still in its first development stages, which leaves numerous opportunities for future work.

First of all the prototype has only been tested on a limited data set, so that the generated synthesis documents could be easily checked. However the synthesis concept aims at helping scientists to get a grasp on large volumes of data. Scale-up experiments have to be conducted to assess the usability of the system on larger data sets, regarding both the number of records and number of parameters.

Secondly the prototype has its deficiencies, including the small number of available components and task models, some limits to the query expressiveness and interactions with the synthesis documents, and a few performance flaws. The first step towards improving the prototype will be to solve these issues. The second step will be to expand on the existing system. For instance we are studying the inclusion of statistical analysis tools in the context of a synthesis document, through the call to R functions. We are also currently planning on adding a new task to explore the distribution of a data set.

Moreover some stages of the process itself are quite complex ; for instance, the organisation of relevant items on the documentary grid is a problem which has elements in common with the classic Operational Research "bin packing" problem. However it uses imbricated boxes of variable sizes, which makes its resolution very difficult. Future works around these complex algorithms have to be considered.

Our prototype also lacks a system to evaluate the quality of the synthesis document, with regards to the structured query. It should be a situational relevance evaluation, which goes beyond classic IR measures. The definition of such a quality measure, taking into account the position of items on the grid, is still in its first stages.

**Acknowledgments.**
J. Bourbeillon is funded by the Minstère de l'Education Nationale, de la Recherche et des Technologies (French Ministry of Research and Technology). Part of this research was also supported by the "Bio-Informatique" inter-EPST program and "La Ligue Contre le Cancer", Savoie committee. The authors would like to thank Dr. Joëlle Simony-Lafontaine who conducted the anatomopathological evaluations of samples, all team members who participated in the user study and the anonymous reviewers whose comments contributed to the improvement of this paper.